\documentclass[12pt]{iopart}

\usepackage{cite}

\usepackage{graphicx}

\begin{document}

\title[Small-energy series for non-symmetric potentials]
{Small-energy series for one-dimensional quantum-mechanical models
with non-symmetric potentials }

\author{Paolo Amore\dag \ and Francisco M Fern\'andez
\footnote[2]{Corresponding author}}

\address{\dag\ Facultad de Ciencias, Universidad de Colima, Bernal D\'iaz del
Castillo 340, Colima, Colima, Mexico}\ead{paolo.amore@gmail.com}

\address{\ddag\ INIFTA (UNLP, CCT La Plata-CONICET), Divisi\'on Qu\'imica Te\'orica,
Blvd. 113 S/N,  Sucursal 4, Casilla de Correo 16, 1900 La Plata,
Argentina}\ead{fernande@quimica.unlp.edu.ar}

\maketitle

\begin{abstract}
We generalize a recently proposed small-energy expansion for
one-dimensional quantum-mechanical models. The original approach
was devised to treat symmetric potentials and here we show how to
extend it to non-symmetric ones. Present approach is based on
matching the logarithmic derivatives for the left and right
solutions to the Schrödinger equation at the origin (or any other
point chosen conveniently) . As in the original method, each
logarithmic derivative can be expanded in a small-energy series by
straightforward perturbation theory. We test the new approach on
four simple models, one of which is not exactly solvable. The
perturbation expansion converges in all the illustrative examples
so that one obtains the ground-state energy with an accuracy
determined by the number of available perturbation corrections.

\end{abstract}

\section{Introduction}

\label{sec:intro}

In a recent paper Bender and Jones\cite{BJ14} proposed a convergent
perturbation series for the calculation of the eigenvalues of the
Schr\"{o}dinger equation in one dimension. The approach consists of the
expansion of the eigenfunction as a power series of the energy $E$ itself
and the construction of a function $f(E)$ that vanishes when $E=0$ and
increases monotonously till $f(E_{0})=1$, where $E_{0}$ is the lowest
eigenvalue. This strategy is based on the fact that the eigenfunction $\psi
(x,E)$ satisfies $\psi _{0}(0,E_{0})=0$ in the case of symmetric potentials $%
V(-x)=V(x)$. The authors also showed how to extend the approach for the
treatment of parity-time invariant complex potentials $V(-x)^{*}=V(x)$. The
method is not restricted to the ground state; the zeros of the Pad\'{e}
approximants for the small-energy expansion of $f(E)-1$ are estimates of the
energies of the excited states\cite{BJ14}.

The purpose of this paper is to extend the approach proposed by Bender and
Jones to non-symmetric potentials. In section~\ref{sec:method} we outline
the method of Bender and Jones but focus on the logarithmic derivative of
the eigenfunction instead of on the eigenfunction itself. In section~\ref
{sec:finite_symmetric} we briefly consider an exactly solvable symmetric
potential. However, instead of choosing an example that supports an infinite
number of eigenvalues like those of Bender and Jones we concentrate on a
finite well. In section~\ref{sec:finite_non-sym} we discuss a finite
non-symmetric well and develop the extension of the method of Bender and
Jones in terms of the logarithmic derivative of the wavefunction. In section~%
\ref{sec:infinite_wells} we illustrate the application of the the
generalized approach on three non-symmetric infinite wells, one of which is
not exactly solvable. Finally, in section~\ref{sec:conclusions} we summarize
the main results and draw conclusions.

\section{The method of Bender and Jones}

\label{sec:method}

Consider the Schr\"{o}dinger equation
\begin{equation}
\psi ^{\prime \prime }(x)=[V(x)-E]\psi (x)  \label{eq:Schro}
\end{equation}
where $\psi (x)$ vanishes at $\pm \infty $. The method applies to symmetric
potentials $V(-x)=V(x)$ that are bounded from below and, without loss of
generality, we assume that $V(x)\geq V(0)=0$.

The method of Bender and Jones\cite{BJ14} is based on the small-energy
expansion for $\psi (x)$
\begin{equation}
\psi (x,E)=\sum_{j=0}^{\infty }\psi _{j}(x)E^{j}  \label{eq:psi_series}
\end{equation}
For convenience, in what follows we consider the alternative expansion of
the logarithmic derivative
\begin{equation}
L(x,E)=\frac{\psi ^{\prime }(x,E)}{\psi (x,E)}  \label{eq:L(x)}
\end{equation}
in the form
\begin{equation}
L(x,E)=\sum_{j=0}^{\infty }L_{j}(x)E^{j}  \label{eq:L_E-expansion}
\end{equation}

Because of the symmetry of the potential the solutions to the equation (\ref
{eq:Schro}) are either even or odd. In particular, the ground state $\psi
(x,E_{0})$ is even so that $\psi ^{\prime }(0,E_{0})=0$ and $L(0,E)$
vanishes when $E$ is the ground-state energy $E_{0}$. Therefore, the
function
\begin{equation}
f(E)=1-\frac{L(0,E)}{L_{0}(0)}  \label{eq:f(E)}
\end{equation}
satisfies $f(0)=0$ and $f(E_{0})=1$. Besides, $f(E)$ can be expanded as\cite
{BJ14}
\begin{equation}
f(E)=\sum_{j=1}^{\infty }c_{j}E^{j}  \label{eq:f(E)_E-expansion}
\end{equation}
Since $\psi (x,E)$ is odd when $E$ is the energy of the first-excited state $%
E_{1}$, then $\psi (0,E_{1})=0$ and $E=E_{1}$ is a pole of $f(E)$.
Consequently, the radius of convergence of (\ref{eq:f(E)_E-expansion})
cannot be greater than $E_{1}$.

Note that we can also obtain $E_{0}$ as a root of $L(0,E)=0$ that can be
obtained approximately from the small-energy series
\begin{equation}
L(0,E)=\sum_{j=0}^{\infty }L_{j}(0)E^{j}  \label{eq:L(0)_series}
\end{equation}

\section{Finite symmetric well}

\label{sec:finite_symmetric}

Bender and Jones studied several symmetric wells that are unbounded from
above and therefore support an infinite number of bound states. Although the
aim of this paper is the application of the small-energy expansion to
non-symmetric potentials we first consider symmetric wells that are bounded
from below and above. Without loss of generality we assume that $0\leq
V(x)\leq V_{R}$. It is well-known that such a potential supports a bound
state no matter how small the well depth $V_{R}$. If there is only one bound
state, then one expects the pole of $f(E)$ closest to the origin in the
complex $E$ plane to be a resonance. This is the main reason for discussing
such symmetric well here.

The simplest exactly-solvable model is given by
\begin{equation}
V(x)=\left\{
\begin{array}{c}
0,\;|x|<1 \\
V_{R}>0,\;|x|>1
\end{array}
\right.  \label{eq:V(x)_well_sym}
\end{equation}
A straightforward calculation yields the logarithmic derivative at the
origin
\begin{equation}
L(0,E)=\frac{\sqrt{E}\left( \sqrt{E}\sin {\left( \sqrt{E}\right) }-\sqrt{%
V_{R}-E}\cos {\left( \sqrt{E}\right) }\right) }{\sqrt{E}\cos {\left( \sqrt{E}%
\right) }+\sqrt{V_{R}-E}\sin {\left( \sqrt{E}\right) }}  \label{eq:L(0)_sym}
\end{equation}
that can be expanded in the $E$-series
\begin{eqnarray}
L(0,E) &=&-\frac{\sqrt{V_{R}}}{\sqrt{V_{R}}+1}+\frac{\left(
2V_{R}^{3/2}+6V_{R}+6\sqrt{V_{R}}+3\right) E}{6\sqrt{V_{R}}\left( \sqrt{V_{R}%
}+1\right) ^{2}}  \nonumber \\
&&+\frac{\left(
8V_{R}{}^{3}+48V_{R}^{5/2}+120V_{R}{}^{2}+180V_{R}^{3/2}+180V_{R}+135\sqrt{%
V_{R}}+45\right) E^{2}}{360V_{R}^{3/2}\left( \sqrt{V_{R}}+1\right) ^{3}}
\nonumber \\
&&+\ldots  \label{eq:L(0)_sym_series}
\end{eqnarray}
Its radius of convergence is determined by the root $E_{r}$ of $\sqrt{E_{r}}%
\cos {\left( \sqrt{E_{r}}\right) }+\sqrt{V_{R}-E_{r}}\sin {\left( \sqrt{E_{r}%
}\right) }=0$ with the smallest absolute value. As an illustrative example
we choose $V_{R}=1$ that supports only one bound state. In this case we have
$E_{r}=1.222635745-0.5925040566i$. Since $E_{0}<1<|E_{r}|$ the perturbation
expansion will enable us to obtain the lowest eigenvalue $E_{0}$ with
increasing accuracy. For example, table~\ref{tab:E0_Well_sym} shows the rate
of convergence of the approximate eigenvalue estimated from the expansion (%
\ref{eq:L(0)_sym_series}) for $V_{R}=1$.

\section{Finite non-symmetric well}

\label{sec:finite_non-sym}

As argued in the introduction, the aim of this paper is the extension of the
method proposed by Bender and Jones to non-symmetric potentials that are
bounded from below. As before, without loss of generality we assume that $%
V(x)\geq V(0)=0$. For a given value of $E$ we construct the logarithmic
derivatives $L_{L}(x,E)$ and $L_{R}(x,E)$ from the left and right solutions $%
\psi ^{(L)}(x,E)$ and $\psi ^{(R)}(x,E)$ to the differential equation (\ref
{eq:Schro}) that vanish when $x\rightarrow -\infty $ and x$\rightarrow
\infty $, respectively. The continuity of the eigenfunction and its first
derivative at $x=0$ requires that the curves $L_{L}(0,E)$ and $L_{R}(0,E)$
intersect at each of the eigenvalues $E=E_{n}$. Obviously, we can obtain $E$%
-power series for both $L_{L}(0,E)$ and $L_{R}(0,E)$ in the way proposed by
Bender and Jones; that is to say, we simply apply the method twice, one to
the left of $x=0$ and one to the right. The intersection of the two
small-energy expansions thus derived provide an estimate of the lowest
eigenvalue $E_{0}$.

In order to illustrate this approach we choose the exactly solvable
non-symmetric finite well
\begin{equation}
V(x)=\left\{
\begin{array}{c}
V_{L}>0,\;x<-1 \\
0,\;|x|<1 \\
V_{R}>0,\;x>1
\end{array}
\right.  \label{eq:V(x)_well_non-sym}
\end{equation}
The exact logarithmic derivatives at the origin
\begin{eqnarray}
L_{L}(0,E) &=&\frac{\sqrt{E}\left( \sqrt{V_{L}-E}\cos {\left( \sqrt{E}%
\right) }-\sqrt{E}\sin {\left( \sqrt{E}\right) }\right) }{\sqrt{E}\cos {%
\left( \sqrt{E}\right) }+\sqrt{V_{L}-E}\sin {\left( \sqrt{E}\right) }}
\nonumber \\
L_{R}(0,E) &=&\frac{\sqrt{E}\left( \sqrt{E}\sin {\left( \sqrt{E}\right) }-%
\sqrt{V_{R}-E}\cos {\left( \sqrt{E}\right) }\right) }{\sqrt{E}\cos {\left(
\sqrt{E}\right) }+\sqrt{V_{R}-E}\sin {\left( \sqrt{E}\right) }}
\label{eq:LR_LL}
\end{eqnarray}
can be expanded as
\begin{eqnarray}
L_{L}(0,E) &=&\frac{\sqrt{V_{L}}}{\sqrt{V_{L}}+1}-\frac{\left(
2V_{L}^{3/2}+6V_{L}+6\sqrt{V_{L}}+3\right) E}{6\sqrt{V_{L}}\left( \sqrt{V_{L}%
}+1\right) ^{2}}  \nonumber \\
&&-\frac{\left(
8V_{L}{}^{3}+48V_{L}^{5/2}+120V_{L}{}^{2}+180V_{L}^{3/2}+180V_{L}+135\sqrt{%
V_{L}}+45\right) E^{2}}{360V_{L}^{3/2}\left( \sqrt{V_{L}}+1\right) ^{3}}%
+\ldots  \nonumber \\
L_{R}(0,E) &=&-\frac{\sqrt{V_{R}}}{\sqrt{V_{R}}+1}+\frac{\left(
2V_{R}^{3/2}+6V_{R}+6\sqrt{V_{R}}+3\right) E}{6\sqrt{V_{R}}\left( \sqrt{V_{R}%
}+1\right) ^{2}}  \nonumber \\
&&+\frac{\left(
8V_{R}{}^{3}+48V_{R}^{5/2}+120V_{R}{}^{2}+180V_{R}^{3/2}+180V_{R}+135\sqrt{%
V_{R}}+45\right) E^{2}}{360V_{R}^{3/2}\left( \sqrt{V_{R}}+1\right) ^{3}}
\nonumber \\
&&+\ldots  \label{eq:LL_LR_series}
\end{eqnarray}

\begin{figure}[H]
\begin{center}
\bigskip\bigskip\bigskip \includegraphics[width=9cm]{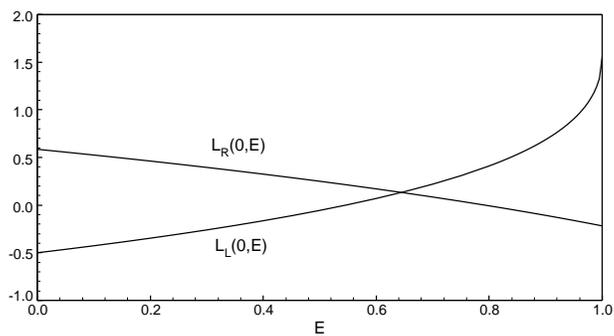}
\end{center}
\caption{$L_R(E,0)$ and $L_L(E,0)$ for the non-symmetric well (\ref
{eq:V(x)_well_non-sym}) with $V_L=2$ and $V_R=1$}
\label{fig:LR-LL}
\end{figure}

Figure~\ref{fig:LR-LL} shows $L_{L}(0,E)$ and $L_{R}(0,E)$ when $V_{L}=2$
and $V_{R}=1$. These curves intersect at the ground-state energy as argued
above. For such values of the potential parameters there is just one bound
state. Table~\ref{tab:E0_Well_asym} shows the rate of convergence of the
estimated lowest eigenvalue obtained from the intersection of the series (%
\ref{eq:LL_LR_series}) for increasing truncation order. In this case the
radii of convergence of each expansion is determined by $%
E_{rL}=2.125364032-0.3468294596i$ and $E_{rR}=1.222635745-0.5925040566i$ and
the approach is successful because both series converge at $%
E=E_{0}<|E_{rR}|<|E_{rL}|$.

\section{Infinite wells}

\label{sec:infinite_wells}

As an extension of the model with the linear symmetric potential $V(x)=|x|$
discussed by Bender and Jones we consider the non-symmetric version
\begin{equation}
V(x)=\left\{
\begin{array}{c}
-a_{L}x,\;x<0 \\
a_{R}x,\;x>0
\end{array}
\right.  \label{eq:V_linear}
\end{equation}
where $a_{L},\,a_{R}>0$. In this case we have
\begin{eqnarray}
L_{L}(0,E) &=&-\frac{a_{L}^{1/3}A_{i}^{\prime }\left( -E/a_{L}^{2/3}\right)
}{A_{i}\left( -E/a_{L}^{2/3}\right) }  \nonumber \\
L_{R}(0,E) &=&\frac{a_{R}^{1/3}A_{i}^{\prime }\left( E/a_{R}^{2/3}\right) }{%
A_{i}\left( E/a_{R}^{2/3}\right) }  \label{eq:L(0)-linear}
\end{eqnarray}
where $A_{i}(z)$ is one of the Airy functions\cite{GR07}.

In order to carry out a sample calculation we choose $a_{R}=1$ and $a_{L}=2$
and obtain the series
\begin{eqnarray}
L_{L}(0) &=&0.9184964715-0.4218178838\,E-0.05628088100\,E^{2}  \nonumber \\
&&-0.01242379097\,E^{3}-0.003082268481\,E^{4}+\ldots  \nonumber \\
L_{R}(0) &=&-0.7290111325+0.5314572310\,E+0.1125617620\,E^{2}  \nonumber \\
&&+0.03944307773\,E^{3}+0.01553365974\,E^{4}+\ldots
\label{eq:L(0)_linear_series}
\end{eqnarray}
The singularities of $L_{L}(0,E)$ and $L_{R}(0,E)$ for this particular
choice of potential parameters appear at $E=3.711514163$ and $E=2.338107410$%
, respectively. Since both are larger than $E_{0}$ the perturbation
expansions (\ref{eq:L(0)_linear_series}) are suitable for the calculation of
this eigenvalue. Table~\ref{tab:E0_Well_linear} shows the rate of
convergence of the approach for the lowest eigenvalue $E_{0}$.

As a non-symmetric extension of the harmonic oscillator discussed by Bender
and Jones we consider the potential
\begin{equation}
V(x)=\left\{
\begin{array}{c}
a_{L}x^{2},\;x<0 \\
a_{R}x^{2},\;x>0
\end{array}
\right.  \label{eq:V_quadratic}
\end{equation}
where $a_{L},\,a_{R}>0$. In this case we have
\begin{eqnarray}
L_{L}(0) &=&\sqrt{2}a_{L}^{1/4}\frac{D_{(E+\sqrt{a_{L}})/(2\sqrt{a_{L}})}(0)%
}{D_{(E-\sqrt{a_{L}})/(2\sqrt{a_{L}})}(0)}  \nonumber \\
L_{R}(0) &=&-\sqrt{2}a_{R}^{1/4}\frac{D_{(E+\sqrt{a_{R}})/(2\sqrt{a_{R}})}(0)%
}{D_{(E-\sqrt{a_{R}})/(2\sqrt{a_{R}})}(0)}  \label{eq:L(0)_quadratic}
\end{eqnarray}
where $D_{\nu }(z)$ is the parabolic cylinder function\cite{GR07}.

For $a_{L}=2$ and $a_{R}=1$ we have the small-energy expansions
\begin{eqnarray}
L_{L}(0)
&=&0.8038781325-0.4464420544\,E-0.06011306588\,E^{2}-0.01251356963\,E^{3}
\nonumber \\
&&-0.002831976530\,E^{4}+\ldots  \nonumber \\
L_{R}(0)
&=&-0.6759782395+0.5309120676\,E+0.1010977236\,E^{2}+0.02976245185\,E^{3}
\nonumber \\
&&+0.009525595408\,E^{4}+\ldots  \label{eq:L(0)_series_quadratic}
\end{eqnarray}
for the left and right solutions, respectively. Table~\ref
{tab:E0_Well_quadratic} shows the convergence of the lowest eigenvalue of
the non-symetric quadratic well (\ref{eq:V_quadratic}) estimated from the
truncated small-energy series (\ref{eq:L(0)_series_quadratic}).

Finally, we consider the Schr\"{o}dinger equation (\ref{eq:Schro}) with the
non-symmetric anharmonic potential
\begin{equation}
V(x)=x^{4}+\lambda x^{3}  \label{eq:V_anharmonic}
\end{equation}
that is not exactly solvable. Note that the minimum of the
potential-energy function $V_{min}=V(x_{min})$ is not located at
the origin but at $x_{min}=-3\lambda/4$ and that
$V(0)=0>V_{min}=-27\lambda^4/256$ does not agree with the
assumption made above. However, that arbitrary assumption was made
for simplicity and is unnecessary for the application of the
approach.

In this case we do not attempt to calculate the small-energy
series for the left and right logarithmic derivatives but we can
obtain $L_{L}(0,E)$ and $L_{R}(0,E)$ quite accurately by means of
a variant of the Riccati-Pad\'{e} method (RPM)\cite{FT96}.

The logarithmic derivative (\ref{eq:L(x)}) can be expanded in a Taylor
series about $x=0$
\begin{equation}
L(x,E)=\sum_{j=0}^{\infty }g_{j}x^{j}  \label{eq:L_x-series}
\end{equation}
where the coefficients $g_{j}$, $j>0$, depend on both $g_{0}$ and $E$. The
Hankel determinants $H_{D}^{d}(E,g_{0})=\left| g_{i+j+d-1}\right|
_{i,j=1}^{D}$, where $D=2,3,\ldots $ is the determinant dimension and $%
d=0,1,\ldots $, are polynomial functions of $g_{0}$ and $E$. For a given
value of $E$ the RPM condition $H_{D}^{d}(E,g_{0})=0$ yields sequences of
roots $g_{0}^{[D,d]}(E)$, $D=2,3,\ldots $ that converge towards $L_{L}(0,E)$
and $L_{R}(0,E)$. Since the RPM takes into account the left and right
solutions simultaneously we cannot determine which sequence corresponds to
either $L_{L}(0,E)$ or $L_{R}(0,E)$. However, this is not a serious drawback
as we will see in what follows.

\begin{figure}[H]
\begin{center}
\bigskip\bigskip\bigskip \includegraphics[width=9cm]{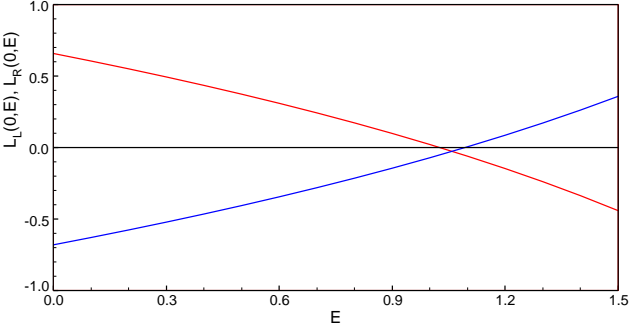}
\end{center}
\caption{$L_R(E,0)$ and $L_L(E,0)$ for the non-symmetric
anharmonic oscillator (\ref{eq:V_anharmonic})}
\label{fig:LR-LL-X4X3}
\end{figure}

For concreteness we restrict ourselves to $\lambda =0.1$ that is small and
we therefore expect $L_{L}(0,E_{0})=L_{R}(0,E_{0})=L(0,E_{0})$ to be close
to zero. Figure~\ref{fig:LR-LL-X4X3} shows that the left and right
logarithmic derivatives approach each other as $E$ increases from $E=0$ and
intersect at $E_{0}$ as expected. In this straightforward application of the
RPM we simply chose $d=0$ and $2\leq D\leq 15$.

We can also calculate the value of $E_{0}$ quite accurately by means of the
standard RPM that is based on pairs of Hankel determinants $%
H_{D}^{d,e}(E,g_{0})=\left| g_{2i+2j+2d-2}\right| _{i,j=1}^{D}$ and $%
H_{D}^{d,o}(E,g_{0})=\left| g_{2i+2j+2d-1}\right| _{i,j=1}^{D}$\cite{FT96}.
In this case, sequences of roots of the set of nonlinear equations $%
H_{D}^{d,e}(E,g_{0})=0$ and $H_{D}^{d,o}(E,g_{0})=0$ converge towards $E_{0}$
and $L(0,E_{0})$. For $\lambda =0.1$ we obtain
\begin{eqnarray}
E_{0} &=&1.0590028460380260258  \nonumber \\
L_{L}(0,E_{0}) &=&L_{R}(0,E_{0})=-0.02652946094577843397
\label{eq:E0_LL_LR_x4x3}
\end{eqnarray}
that agree with the intersection shown in Figure~\ref{fig:LR-LL-X4X3}. Here
we chose the same values of $D$ and $d$ indicated previously.

\section{Conclusions}

\label{sec:conclusions}

In this paper we generalized the method of Bender and Jones so that it can
be applied to non-symmetric potentials. Present approach consists of
matching the logarithmic derivatives of the left and right solutions at the
origin. The matching procedure itself is well known but here it is combined
with the original idea of the small-energy series proposed by those authors.
The series are convergent as in the case of the symmetric potentials studied
earlier. It is worth noting that the same procedure applies to symmetric
potentials but in this case it is not necessary to carry out both
calculations because the two curves $L_{L}(0,E)$ and $L_{R}(0,E)$ are
symmetric with respect to the $E$ axis.

As illustrative examples we explicitly considered three exactly
solvable models and a nontrivial anharmonic oscillator. In the
latter case we did not derive the small-energy series and
restricted ourselves to the accurate calculation of the two
logarithmic derivatives at origin by means of the RPM. In this way
we showed that the two curves intersect at the ground-state energy
that we also calculated accurately by means of the RPM. For the
treatment of one-dimensional or separable problems the RPM is by
far preferable to the perturbation approach but the latter may
hopefully be applied to non-separable systems. For the time being
we do not know how to apply this perturbation approach to
multidimensional problems, but it is likely that one should have
to resort to some kind of matching procedure like the one
illustrated in this paper.

\ack F. M. Fern\'andez would like to thank the University of
Colima for financial support and hospitality

\begin{table}[H]
\caption{Eigenvalue of the symmetric well (\ref{eq:V(x)_well_sym}) of depth $%
V_R=1$ estimated by means of the expansion (\ref{eq:L(0)_sym_series}) of
order $n$. The exact result is $E_0=0.5462468341$.}
\label{tab:E0_Well_sym}
\begin{center}
\par
\begin{tabular}{rr}
$n$ & \multicolumn{1}{c}{$E_0$} \\
4 & 0.5855444198 \\
8 & 0.5516251660 \\
12 & 0.5472622152 \\
16 & 0.5464638914 \\
20 & 0.5462964250 \\
24 & 0.5462586560 \\
28 & 0.5462497386 \\
32 & 0.5462475642 \\
36 & 0.5462470210 \\
40 & 0.5462468826 \\
44 & 0.5462468468 \\
48 & 0.5462468375 \\
52 & 0.5462468350 \\
56 & 0.5462468344 \\
60 & 0.5462468343 \\
64 & 0.5462468341
\end{tabular}
\end{center}
\end{table}

\begin{table}[H]
\caption{Eigenvalue of the non-symmetric well (\ref{eq:V(x)_well_non-sym})
with $V_L=2$ and $V_R=1$ estimated by means of the expansions of $L_R(0,E)$
and $L_L(0,E)$. The exact result is $E_0=0.6446113612$.}
\label{tab:E0_Well_asym}
\begin{center}
\par
\begin{tabular}{rr}
$n$ & \multicolumn{1}{c}{$E_0$} \\
4 & 0.8367722372 \\
8 & 0.6634864161 \\
12 & 0.6487132635 \\
16 & 0.6457463863 \\
20 & 0.6449609259 \\
24 & 0.6447254502 \\
28 & 0.6446499927 \\
32 & 0.6446247871 \\
36 & 0.6446161202 \\
40 & 0.6446130747 \\
44 & 0.6446119860 \\
48 & 0.6446115915 \\
52 & 0.6446114469 \\
56 & 0.6446113933 \\
60 & 0.6446113734 \\
64 & 0.6446113658 \\
68 & 0.6446113630 \\
72 & 0.6446113618 \\
76 & 0.6446113615 \\
80 & 0.6446113614 \\
84 & 0.6446113614 \\
88 & 0.6446113612
\end{tabular}
\end{center}
\end{table}

\begin{table}[H]
\caption{Eigenvalue of the non-symmetric linear well (\ref{eq:V_linear})
with $a_L=2$ and $a_R=1$ estimated by means of the expansions of $L_R(0,E)$
and $L_L(0,E)$. The exact result is $E_0=1.250207832$.}
\label{tab:E0_Well_linear}
\begin{center}
\par
\begin{tabular}{rr}
$n$ & \multicolumn{1}{c}{$E_0$} \\
2 & 1.387352237 \\
4 & 1.275507151 \\
6 & 1.256485215 \\
8 & 1.251913598 \\
10 & 1.250686548 \\
12 & 1.250343776 \\
14 & 1.250246604 \\
16 & 1.250218907 \\
18 & 1.250210997 \\
20 & 1.250208737 \\
22 & 1.250208091 \\
24 & 1.250207905 \\
26 & 1.250207854 \\
28 & 1.250207837 \\
30 & 1.250207833 \\
32 & 1.250207832
\end{tabular}
\end{center}
\end{table}

\begin{table}[H]
\caption{Eigenvalue of the non-symmetric quadratic well (\ref{eq:V_quadratic}%
) with $a_L=2$ and $a_R=1$ estimated by means of the expansions of $L_R(0,E)$
and $L_L(0,E)$. The exact result is $E_0=1.176933152$.}
\label{tab:E0_Well_quadratic}
\begin{center}
\par
\begin{tabular}{rr}
$n$ & \multicolumn{1}{c}{$E_0$} \\
2 & 1.254541164 \\
4 & 1.185370810 \\
6 & 1.178091246 \\
8 & 1.177102981 \\
10 & 1.176958699 \\
12 & 1.176937027 \\
14 & 1.176933737 \\
16 & 1.176933265 \\
18 & 1.176933166
\end{tabular}
\end{center}
\end{table}

\end{document}